\documentclass{article}
\usepackage{epsfig}

\def\be{\begin{equation}}                                                                         
\def\ee{\end{equation}}                                                                         
\def\bea{\begin{eqnarray}}                                                                         
\def\eea{\end{eqnarray}}                                                                         
                                                                         
\begin{document}                                                                         
\title{Ultrahigh energy neutrino physics}                                           
\author{J.~Kwiecinski$^{a}$, A.D.~Martin$^b$ and A.M.~Stasto$^{a}$ \\                                                                 
\small $^a$ H.~Niewodniczanski Institute of Nuclear Physics, 
ul.~Radzikowskiego 152,  Krakow, Poland \\                                              
\small $^b$ Department of Physics, University of Durham, Durham, DH1 3LE }                                                                  
\maketitle
\begin{abstract}
We discuss a problem concerning the ultrahigh energy neutrino propagation
through the Earth. We present  calculation of the neutrino-nucleon cross section
at high energies, based on the unifed evolution equation at small $x$.
 We also show the solution of the transport equation for different neutrino fluxes
originating from active galactic nuclei, gamma ray bursts and top-down model.
\end{abstract}        
\section{Introduction}
The ultrahigh energy neutrino physics is an interdisciplinary subject which can
address many vital problems in astrophysics, particle physics and geophysics, for a review see \cite{GQRS2,GQRS1}.
Neutrinos can traverse large distances without being disturbed and thus give
us important information about distant astronomical objects like active galactic nuclei or gamma
ray bursts. The ultrahigh energy neutrinos can have energies up to $10^{12} \; \rm GeV$,
much larger than currently accessible at present colliders, like HERA. At these ultrahigh energies
the structure of nucleon is probed at very small values of Bjorken $x$. It means
that we are possibly entering a region in which partons have large density.
Therefore detailed knowledge of parton distribution function at very small
$x$ is vital to estimate the neutrino-nucleon cross section.
This cross section rises strongly with  energy and therefore neutrino propagation
through the Earth can be affected by the increased interaction with matter at these ultrahigh energies.
Thus it is necessary to consider the effect of attenuation of high energy neutrinos
while traversing the Earth. This phenomenon could be used as a possible method of the "Earth
tomography" by neutrinos. All these topics will be studied at large neutrino telescopes
like AMANDA, NESTOR and ANTARES (for review see \cite{HALZEN}).

In this paper we present a method of calculation the neutrino-nucleon cross section using
gluon and quark distributions obtained from
a unifed evolution equation for small $x$ \cite{KMS}. This equation embodies both DGLAP and BFKL evolutions
on equal footing as well as important subleading effects via consistency constraint \cite{AGS}.
This constraint limits the virtualities of the exchanged gluon momenta
in the ladder to their transverse components.
We then extrapolate the results to ultrahigh energies and compare the predictions
with the other, based on standard global parton analysis \cite{GQRS2},\cite{GKR}.

We use the resulting cross section as an input to the transport equation \cite{NIK} for the
neutrino flux penetrating the Earth. We solve this equation for different incident angles 
as well as different input neutrino
fluxes originating from active galactic nuclei, gamma ray bursts and a sample top-down model. \\

The details of the  calculation has been presented elsewhere \cite{KMS2}. \\

\section{ Sources of ultrahigh energy neutrinos }

Neutrinos have the advantage that they are weakly interacting with matter
and they are hardly absorbed when travelling along large distances.
On the other hand cosmic rays are being absorbed through the following interactions:
\begin{itemize}
\item  pair production: $\gamma_{CR} + \gamma_{BR} \rightarrow e^+ e^-$ 
\item inverse Compton scattering: $e_{CR} + \gamma_{BR} \rightarrow e + \gamma$
\item photoproduction of pion: $p_{CR} + \gamma_{BR} \rightarrow p + N \pi$
\item  nuclei fragmentation by photo-pion interactions
\end{itemize}
These reactions cause that the spectrum of the cosmic rays has the GZK cutoff \cite{GZK}
around $10^{19} \rm eV$.  Therefore neutrinos are the best candidates for supplying information
about distant objects. The possible sources of highly energetic neutrinos are:
\begin{itemize}
\item active galactic nuclei
\item gamma ray bursts
\item top-down models
\end{itemize}
{\it Active galactic nuclei} are the most powerful sources of radiation in the whole Universe.
The engine of the AGN is the supermassive black hole with mass bilion times larger
than the mass of the sun. Surrounding  the black hole is the accretion disc usually
accompanied by two jets. The spectrum of the emitted photons spreads from radio waves 
to TeV energies. The jets are the sources of the most energetic gamma rays, since
 the particles (electrons and possibly protons) are accelerated  in blobs along the jets
with Lorenz factor $\gamma \sim 10$. Electrons loose energy via synchrotron radiation
thus producing very energetic photons. If protons are also accelerated then they
can interact with the ambient photons producing pions. Consequently strong flux of
neutrinos will be produced \cite{AGN1,AGN2,AGN3}.\\

{\it Gamma ray bursts} are also very probable sources of neutrinos. Although the underlying
event of GRB's is not entirely known the present knowledge is consistent  that the bursts
are produced as a relativistically expanding fireball which intitial radius was around $100 \rm km$.
The original state which is opaque to light expands in a relativistic shock with $\gamma \sim 300$
to the point where it becomes optically thin and produces intensive gamma ray spectrum.
As in the case of AGN, the acceleration of protons can result in the production of neutrinos
because they will photoproduce pions and in the end neutrinos will emerge \cite{GRB}. \\

{\it Top-down models} are the most speculative scenarios of producing neutrino fluxes
at ultrahigh energies. In these models one assumes that the particles are not accelerated
but are rather produced as a result of the decay of the supermassive particles
$M_x \sim 10^{14} - 10^{16} \rm GeV$. These particles could emerge as a decay of some topological
defects: superconducting strings or magnetic monopoles. The top-down models
produce quite hard spectrum of neutrinos extending beyond $10^{12} \rm GeV$ and they
were proposed as a possible solution to the GZK cutoff \cite{TDM}. \\

\section{ Neutrino - nucleon cross section }

\begin{figure}[!h]
\centerline{\epsfig{file=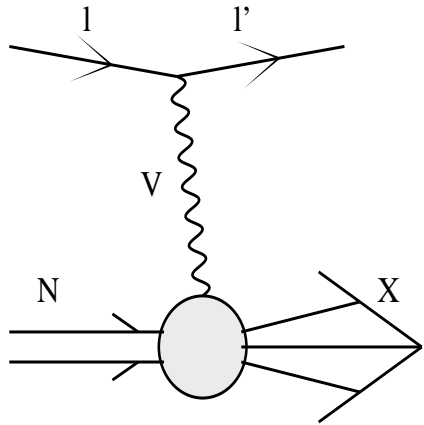,height=6cm,width=6cm}}
\caption{ Deep inelastic scattering. }
\label{fig:fig1}
\end{figure}

The dominant interaction for neutrinos is the neutrino-nucleon interaction.
It has the largest value of the cross section, and is dominating over the interaction
with electrons. There is however one exception:  resonant $W$ production in
$\bar{\nu_{e}} \; e^{-}$ interaction at $E_{\nu} = 6.4 \times 10^5 \; \rm GeV$.
At this energy this process domintates by 2 orders of magnitude over other
contributions. 
The neutrino - nucleon interaction can be visualised in Fig. \ref{fig:fig1}.

\begin{figure}[!h]
\centerline{\epsfig{file=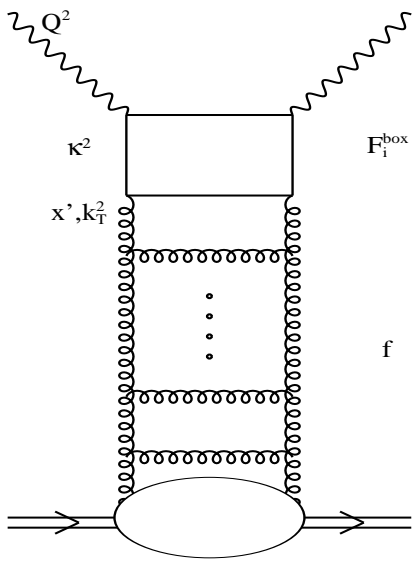,height=12cm,width=6cm}}
\caption{ Diagrammatic representation of the $k_T$-factorization formula.
  At lowest order in $\alpha_S$ the gauge boson-gluon fusion processes,                      
$Vg \rightarrow q\bar{q}$, are given by the quark box shown (together with the                      
crossed box).  The variables $\kappa$, $k$ and $k^\prime$ denote the transverse                      
momenta of the indicated virtual particles.}
\label{fig:fig2}
\end{figure}

Here $l$ is the incoming neutrino of four momentum $k$ and $l'$ is
the outgoing neutrino or charged lepton with four momentum $k'$. $N$ is the target nucleon 
with four momentum $p$
and $X$ is the arbitrary hadronic final state.
This deep inelastic scattering process can be described using two standard variables:
$q^2 = -Q^2 < 0$ is the four momentum squared of the exchanged vector boson,
and  $x = {Q^2 \over 2 p \cdot q}$ is the standard Bjorken scaling variable.
At high energies (up to $10^{12} GeV$) , very small  values of $x$ can be probed because,
\be
x \sim {M_V^2 \over 2 M_N \nu} \sim 10^{-8} 
\ee

where $M_V$ is the mass of the heavy vector boson,  $M_N$  is the nucleon mass and $\nu$ is the 
energy of the exchanged vector boson. These are values of $x$ which are not accessible
at present electron-proton colliders (for example HERA goes down to $10^{-5}$ at fairly low $Q^2$).
Therefore a detailed knowledge of parton distributions at these small values
of $x$ are required.  We propose to use the unifed BFKL/DGLAP evolution equation \cite{KMS}. 
We shall start at first with the pure leading order BFKL equation \cite{BFKL} for the unintegrated gluon distribution
function $f(x,k^2)$ in the followng form:
\be                                        
\label{eq:a1}                                              
f(x, k^2) \; = \; f^{(0)} (x, k^2) \: + \: \overline{\alpha}_S k^2 \int_x^1 \frac{dz}{z}                                         
\int  \frac{dk^{\prime 2}}{k^{\prime 2}} \left \{\frac{f(x/z, k^{\prime 2}) - f(x/z,                                         
k^2)}{| k^{\prime 2} - k^2 |} \: + \: \frac{f(x/z, k^2)}{[4k^{\prime 4} +                                         
k^4]^{\frac{1}{2}}} \right \}                                            
\ee                                        
where $\overline{\alpha}_S = N_c \alpha_S/\pi$ and $k = k_T, k^\prime =                                            
k_T^\prime$ denote the transverse momenta of the gluons, see Fig.~2.  The term in                                            
the integrand containing $f (x/z, k^{\prime 2})$ corresponds to real gluon emission,                                         
whereas the terms involving $f (x/z, k^2)$ represent the virtual contributions                                            
and lead to the Reggeization of the $t$-channel exchanged gluons.  The                                            
inhomogeneous driving term $f^{(0)}$ is the input function will be specified later.      
This is the leading order in $\ln(1/x)$ equation which gives the very well known
intercept for the gluon $\lambda = \bar{\alpha_s} 4 \ln 2$. The next to leading contribution
to the t-channel exchange at high energies has been already calculated \cite{NLO}.
It yields however very large correction to the intercept making it physically unreliable.
It occurred, that the resummation of the subleading effects should be performed in order to get
physically reliable results \cite{SALAM,CCS}.  Different forms of resummation has been already proposed in the literature.
It appears that the imposition of the consistency constraint \cite{AGS} which limits the phase-space available
for the real emission term provides with the partial resummation of the subleading effects.
This constraint arises from the fact  that in the high energy limit the virtualities of the
exchanged momenta are dominated by their transverse parts. By imposition of this constraint one
obtains  nice physical picture in which the subleading effects are resummed by the limitation
of the phase space. The result for the gluon intercept yields reasonable value which 
is stable, i.e. does not become negative.
Second improvement to the equation (\ref{eq:a1})  is the inclusion of the DGLAP terms 
which are important for large values of $x$ and the overall normalisation of the resulting gluon
distribution function. We do also include the quark driving term in eq. \ref{eq:a1}.
The resulting evolution equation for the gluon has the following form:
\bea                                            
\label{eq:a2}                                            
& & f(x, k^2) \; = \; \tilde{f}^{(0)} (x, k^2) + \nonumber \\                                           
& & \nonumber \\               
& & + \: \overline{\alpha}_S (k^2) k^2 \int_x^1 \frac{dz}{z}                                             
\int_{k_0^2} \frac{dk^{\prime 2}}{k^{\prime 2}} \left\{\frac{f                                            
\left( {\displaystyle \frac{x}{z}}, k^{\prime 2} \right) \Theta \left({\displaystyle                                           
\frac{k^2}{z}} - k^{\prime 2}\right) - f \left({\displaystyle \frac{x}{z}}, k^2\right)}                                            
{| k^{\prime 2} - k^2 |} \; + \; \frac{f \left({\displaystyle \frac{x}{z}}, k^2                                           
\right)}{[4k^{\prime 4}+ k^4]^{\frac{1}{2}}} \right \} \\                                            
& & \nonumber \\                                            
& & + \: \overline{\alpha}_S (k^2) \int_x^1 \frac{dz}{z} \left(\frac{z}{6}                                            
P_{gg} (z) - 1 \right ) \int_{k_0^2}^{k^2}  \frac{dk^{\prime 2}}                                            
{k^{\prime 2}} f \left(\frac{x}{z}, k^{\prime 2} \right ) \: + \:                                            
\frac{\alpha_S (k^2)}{2\pi} \int_x^1 dz P_{gq} (z) \Sigma                                             
\left(\frac{x}{z}, k^2 \right ). \nonumber                                         
\eea                                            
We specify the driving term in  the following form                                            
\be                                          
\label{eq:a3}                                          
\tilde{f}^{(0)} (x, k^2) \; = \;  
\frac{\alpha_S (k^2)}{2\pi} \int_x^1 dz P_{gg} (z) \frac{x}{z} g                                            
\left(\frac{x}{z}, k_0^2 \right)                                          
\ee                                          
Let us note that the inhomogenious term has been entirely specifed 
in terms of the standard integrated gluon distribution function
at the infrared cutoff $k_0^2$
and that the BFKL/DGLAP evolution only takes place above this
cutoff. 
The last term is the contribution of the quark distribution to the
gluon evolution,

\be                                         
\label{eq:a6}                                         
\Sigma \; = \; \sum_q \: x (q + \bar{q}) \; = \; \sum_q (S_q + V_q)                                         
\ee                                          
where $S$ and $V$ denote the sea and valence quark momentum distributions.  The                                          
gluon, in turn, helps to drive the sea quark distribution through the $g \rightarrow                                          
q\bar{q}$ transition.  Thus equation (\ref{eq:a2}) has to be solved simultaneously                                          
with an equivalent equation for $\Sigma (x, k^2)$.                                         
We use the $k_T$ factorisation formula \cite{KTFAC}, 
see Fig. \ref{fig:fig2}  as a basis for our evolution equation for the quark
distribution.   
\be                                         
\label{eq:a7}                                         
S_q (x, Q^2) \; = \; \int_x^1 \: \frac{dz}{z} \: \int  \: \frac{dk^2}{k^2} \: S_{\rm                                          
box}^q \: (z, k^2, Q^2) \: f \left  ( \frac{x}{z}, k^2 \right )                                        
\ee                                          
where $S^{\rm box}$ describes the quark box (and crossed-box) contribution shown                                          
in Fig.~2 and can be interpreted as a partonic structure function.
It has been shown (see for example \cite{THORNE}, \cite{KMS}) that the $\ln(1/x)$ effects are also resummed
and play important role in the $k_T$ factorisation prescription.
In order to get the complete set of evolution equations we also have to add quark-quark
splittings (which are small numerically anyway)  and the valence quarks,
which we take from the set of parametrisations.
Thus our complete equation for the singlet quark distribution reads as follows,
\bea                                          
\label{eq:a10}                                          
\Sigma (x, k^2) & = & S_{\rm non-p} (x) \: + \: \sum_q \int_{x}^a                                          
\frac{dz}{z} \: S_q^{\rm box} (z, k^{\prime 2} = 0, k^2) \frac{x}{z} \: g                                          
\left(\frac{x}{z},  k_0^2 \right) \nonumber \\                                          
& & \nonumber \\                                          
& & + \: \sum_q \int_{k_0^2}^\infty \frac{dk^{\prime                                          
2}}{k^{\prime 2}} \int_x^1 \frac{dz}{z} \: S_q^{\rm box} (z, k^{\prime 2},                                          
k^2) f \left (\frac{x}{z}, k^{\prime 2} \right) \: + \: V (x, k^2)  \nonumber  \\                                          
& & \\                                          
& & + \: \int_{k_0^2}^{k^2} \frac{dk^{\prime 2}}{k^{\prime 2}} \:                                          
\frac{\alpha_S  (k^{\prime 2})}{2\pi} \int_x^1 dz \: P_{qq} (z)                                          
S_{uds} \left(\frac{x}{z}, k^{\prime 2} \right ) \nonumber                                          
\eea                                         
where $a = (1 + 4m_q^2/Q^2)^{-1}$ and $V = x (u_v + d_v)$. 
Equations (\ref{eq:a2}) and (\ref{eq:a10})   form a set of coupled integral equations
for the unkown functions $f(x,k^2)$ and $\Sigma(x,k^2)$. We solve them assuming simple
parametric form of the inputs: 
\bea
xg(x,k_0^2) &  = & N(1-x)^{\beta} \nonumber \\
S_{\rm non-p}(x) & = & C_p (1-x)^8 x^{-0.08} 
\label{eq:a11}
\eea

We have used this set of coupled equations to calculate $F_2$ structure function at HERA,                     
and we have therefore fixed the values of the free parameters.
We then used the resulting parton distribution functions in order to calculate the 
neutrino - nucleon cross section at high energies.
We calculate the cross sections from the usual formula: 
\bea                                        
\label{eq:a11}                                        
\frac{d^2 \sigma^{\nu, \overline{\nu}}}{d x d y} & = & \frac{G_F ME}{\pi} \left (                                         
\frac{M_i^2}{Q^2 + M_i^2} \right )^2 \: \left \{ \frac{1 + (1 - y)^2}{2} \: F_2^\nu (x,                                         
Q^2) \right . \nonumber \\                                        
& & \\                                        
& & - \: \left . \frac{y^2}{2} \: F_L^\nu (x, Q^2) \: \underline{+} \: y \left ( 1 -                                         
\frac{y}{2} \right ) \: x F_3^\nu (x, Q^2) \right \} \nonumber                                        
\eea                                        
where $G_F$ is the Fermi coupling constant, $M$ is the proton mass, $E$ is the                                         
laboratory energy of the neutrino and $y = Q^2/xs$.  The mass $M_i$ is either                                         
$M_W$ or $M_Z$ according to whether we are calculating charged current (CC) or                                         
neutral current (NC) neutrino interactions.             
\begin{figure}[!h]
\centerline{\epsfig{file=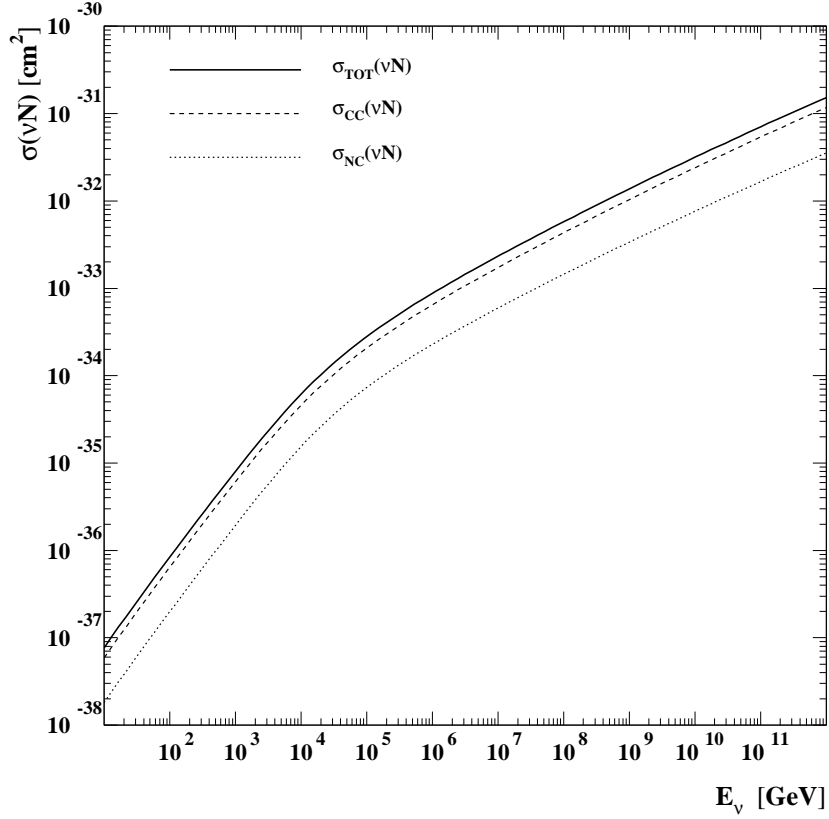,height=12cm,width=12cm}}
\caption{The total $\nu N$  cross section and its decomposition                      
into contributions from charged current and neutral current
 plotted as a function of the laboratory neutrino energy.}
\label{fig:fig3}
\end{figure} 
Functions $F_2^{\nu}$, $F_L^{\nu}$ and $xF_3^{\nu}$ are of course usual structure functions which can be calculated
from the parton distribution functions.\\
In Fig.~3 we show the plot of $\sigma(\nu N)$ as a function of energy
of the neutrino $E_{\nu}$. One observes strong rise, nearly $8$ decades
with increasing energy from $10$ to $10^{12} \rm GeV$. The shape of the curves
up to $10^5 \rm GeV$ is determined by the valence quarks whereas
beyond $10^6 \rm GeV$ everything is driven by the sea quarks.
One also notes  that the charged current contribution is dominating over the neutral
current by factor 3.

\section{Transport equation} 

Neutrinos at ultrahigh energies can be quite strongly attenuated
when traversing the Earth. Apart from standard absorption neutrinos can undergo regeneration
due to neutral current interactions. Charged current interactions remove neutrinos
from the flux, but neutral current interaction cause neutrinos to reappear at lower energies.
 These both effects can be calculated using the transport equation
proposed by \cite{NIK}:
\be                                   
\label{eq:a18}                                   
{dI(E,\tau)\over d\tau} \; = \; - \sigma_{\rm TOT}(E) I(E,\tau) \: + \: \int {dy\over 1-y}                                    
{d\sigma_{\rm NC}(E^{\prime},y) \over dy} I(E^{\prime},\tau)                                   
\ee                                   
where $\sigma_{\rm TOT} = \sigma_{\rm CC} + \sigma_{\rm NC}$ and where $y$                                    
is, as usual, the fractional energy loss such that                                   
\be                                   
\label{eq:a19}                                   
E^{\prime}={E\over 1-y}.                                   
\ee                                  
The variable $\tau$ is the number density of nucleons $n$ integrated along a path of                                    
length $z$ through the Earth                                   
\be                                   
\label{eq:a20}                                   
\tau(z) \; = \; \int_0^z \: dz^\prime \: n (z^\prime).                                   
\ee                 
The number density $n(z)$ is defined as $n(z)=N_A \: \rho (z) $ where $\rho (z)$ is the 
density of Earth along the neutrino path length $z$ and $N_A$ is the Avogadro number.                  
The number of nucleons $\tau$ encountered along the path $z$ depends upon                                    
the nadir angle $\theta$ between the normal to the Earth's surface (passing through the                                    
detector) and the direction of the neutrino beam incident on the detector.  
\begin{figure}[!h]
\centerline{\epsfig{file=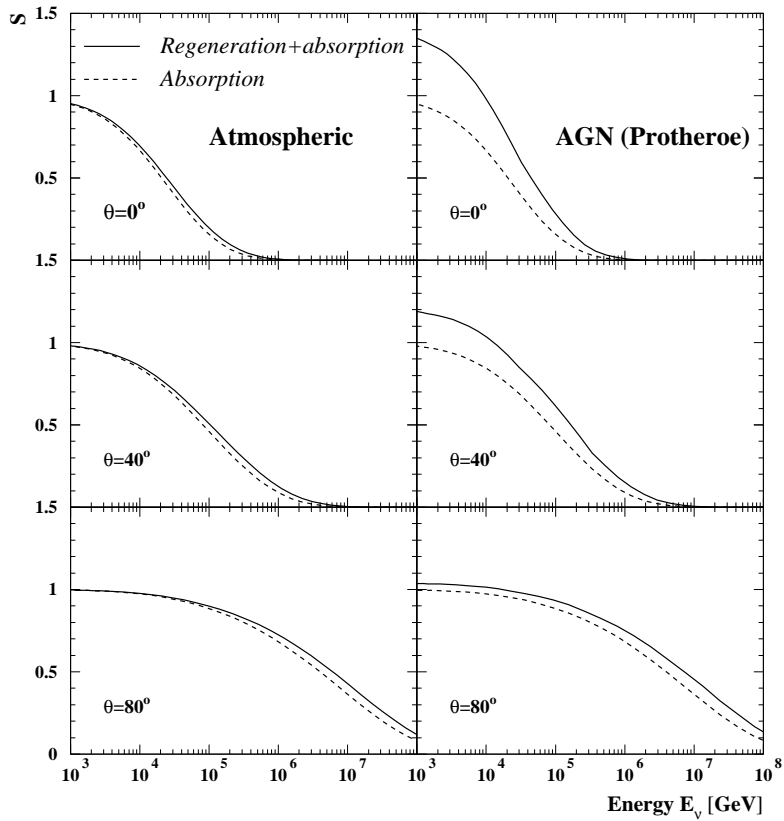,height=12cm,width=12cm}}
\caption{ The shadowing factor $S$ of (\ref{eq:a21}) for two different initial                      
neutrino fluxes incident at three different nadir angles on a detector.  The angle $\theta                      
= 0^\circ$ corresponds to penetration right through the Earth's diameter.  The two                      
curves on each plot show the shadowing factor with and without NC regeneration                      
included.}
\label{fig:fig4}
\end{figure} 

In order to calculate the change of the intensity of the flux with the incident angle
one needs to know the density profile of the Earth. We have used the model by A. Dziewo\'nski
\cite{EARTH}. 
 Using this parametrisation of the Earth density and the cross sections calculated from
the unifed BFKL/DGLAP equation. In Fig.~4 we show the shadowing factor,
\be
S(E,\tau) \; = \; {I(E,\tau) \over I^0(E)}
\label{eq:a21} 
\ee
where $I^0(E) = I(E,\tau)$ is the initial flux at the surface of the Earth.
We present $S$ as a function of the energy and for different incident angles.
The curves exhibit strong suppression for large paths in matter and the energies above $10^6 \rm GeV$.
Also, the curves corresponding to the flux from AGN \cite{AGN1}, show that the regeneration is important
for flat fluxes and large paths. The same effect is nearly negligible in the case of steeply
falling atmospheric spectrum \cite{ATMOS}.
\begin{figure}[!h]
\centerline{\epsfig{file=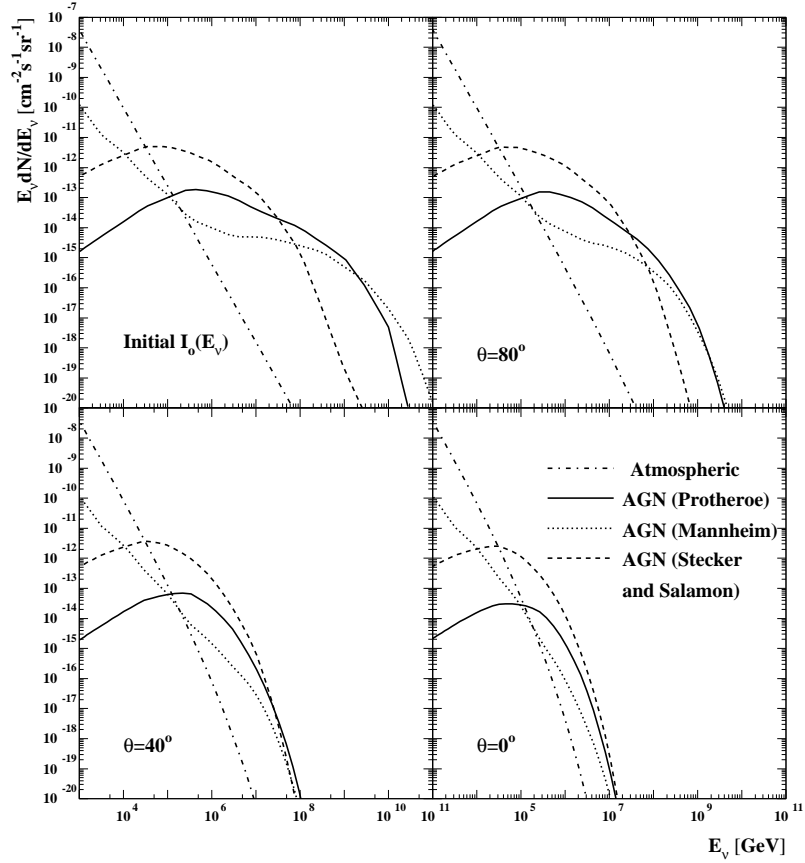,height=12cm,width=12cm}}
\caption{The initial flux $I_0 (E)$ and the flux at the detector $I (E)$ for three                      
different nadir angles corresponding to three models for AGN neutrinos                 
\cite{AGN1,AGN2,AGN3}.  The background atmospheric neutrino flux is also                 
shown.  All the fluxes are given for muon neutrinos. The corresponding fluxes 
from \cite{AGN1,AGN2,AGN3} were given originally for muon neutrinos
and anti-neutrinos, and their value has been divided by factor 2.}
\label{fig:fig5}
\end{figure} 
\begin{figure}[!h]
\centerline{\epsfig{file=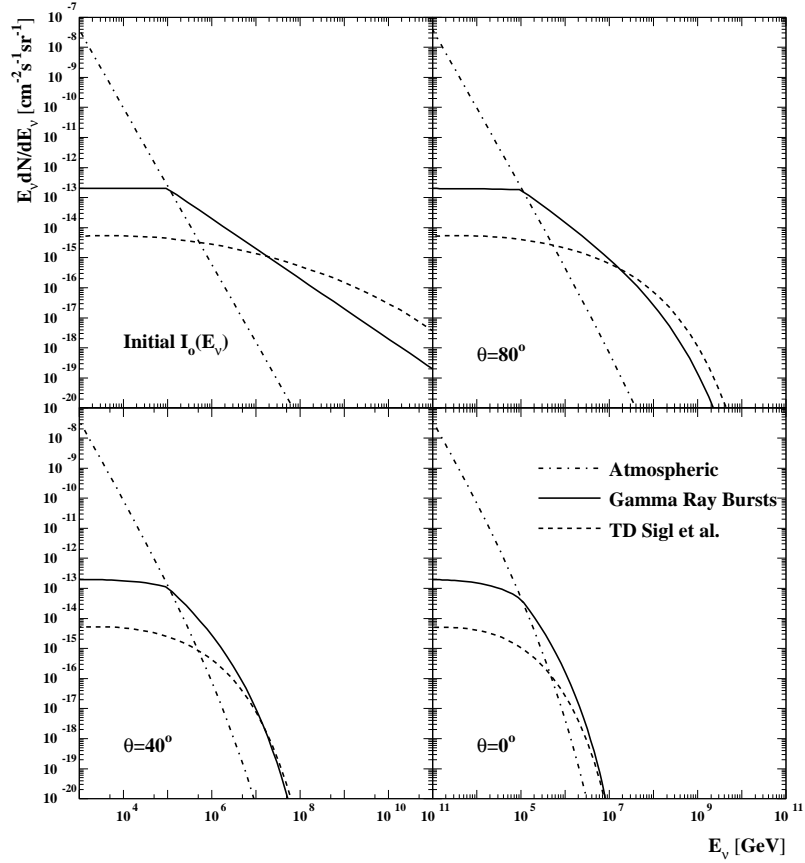,height=12cm,width=12cm}}
\caption{The initial flux $I_0 (E)$ and the flux at the detector $I (E)$ for three                      
different nadir angles corresponding to the model of gamma ray burst               
\cite{GRB} and the top-down model \cite{TDM}.  The background atmospheric neutrino flux is also                 
shown.  All the fluxes are given for muon neutrinos. }
\label{fig:fig5}
\end{figure} 

In Figs.~5 and ~6 we show complete simulation for different incident fluxes corresponding 
to the active galactic nuclei \cite{AGN1,AGN2,AGN3}, gamma ray bursts \cite{GRB}
and a sample top-down model \cite{TDM}. 
Similar effects of strong attenuation are observed for large energies $\sim 10^6 \rm GeV^2$
and large paths. This suggests that one will have to choose suitable angle 
of the observation in order to avoid large muon background from the atmosphere
and yet be able to detect the highly energetic neutrinos which are likely to be absorbed
by the matter in Earth.\\

{\Large \bf Summary} \\

In this paper we have examined the interactions of ultrahigh energy neutrinos with matter.
As the cross section rises with increasing energy the absorption by matter becomes
interestingly large. We have seen that the Earth becomes essentially opaque to ultrahigh
energy neutrinos. We have calculated neutrino-nucleon cross section using the unifed DGLAP/BFKL
evolution equations which treat leading $\ln(Q^2)$ and $\ln(1/x)$ on equal footing.
These equations also resum important subleading effects in $\ln(1/x)$ via imposition of consistency
constraint. We believe that this form of the evolution is most apropriate in the regime
where the parton density is large.
The results for the cross section has been compared with the other based on
the standard global parton analysis. In that way the uncertainty due to parton density extrapolation
has been diminished to $40\%$. We have then used the resulting cross sections
and solved the transport equation for the neutrinos travelling through the Earth. 
We have found that the attenuation is large for high energies , above $10^6 \rm GeV$ and
large paths in matter. We have also found that the regeneration due to neutral current
interactions becomes important for flat spectra (like these originating from active galactic nuclei)
and large paths. We have simulated the penetration through the Earth for different  neutrino fluxes:
active galactic nuclei, gamma ray bursts and top-down models. 
The large attenuation effect reduces substantially the flux for small nadir angles.
There is however a window for observation of AGN fluxes by the $\rm km^3$ detectors.\\

\noindent {\Large \bf Acknowledgements}                \\
                
    This research has been supported in part by the                 
Polish State Committee for Scientific Research (KBN) grant N0~2~P03B~89~13                 
and by the EU Fourth Framework Programme \lq\lq Training and                 
Mobility of Researchers", Network \lq\lq Quantum Chromodynamics and the Deep                 
Structure of Elementary Particles", contract FMRX - CT98 - 0194.                                


\end{document}